\documentclass[hyper]{JHEP} % 10pt is ignored!
\usepackage{epsfig}
\newcommand\fverb{\setbox\pippobox=\hbox\bgroup\verb}
\newcommand\fverbdo{\egroup\medskip\noindent%
            \fbox{\unhbox\pippobox}\ }

\newcommand\fverbit{\egroup\item[\fbox{\unhbox\pippobox}]}
\newbox\pippobox

\title{Giant Magnon in NS5-brane Background}

\preprint{\hepth{0703244}}
\author{J. Kluso\v{n}\\
Dipartimento di Fisica,\\
Universita' \& I.N.F.N. Sezione di Roma 2, Tor Vergata \\
Via della Ricerca Scientifica 1, 00133  Roma,   ITALY\\
E-mail: \email{Josef.Kluson@roma2.infn.it}}
\author{Rashmi R. Nayak \\
Center for Quantum Spacetime (CQUeST),\\
Sagang University, Seoul, 121-742, Korea\\
E-mail: \email{rashmi@sogang.ac.kr}}
\author{Kamal L. Panigrahi\\
Department of Physics\\
Indian Institute of Technology, Guwahati, 781 039, Guwahati, INDIA \\
E-mail: \email{panigrahi@iitg.ernet.in}} \abstract{We study the
giant magnon solutions in the near horizon geometry of the
Neveu-Schwarz (NS) 5-brane background. In conformal gauge, we find
magnon dispersion relation in the large angular momentum $(J)$
limit. We further show that the giant magnon poses uniform
distribution of the angular momentum along the string world-sheet
as in case of $AdS_5\times S^5$ spacetime.} 

\def\bp{\mathbf{p}}

\def\pb  #1{\left\{#1\right\}}

\newcommand{\bK}{\mathbf{K}}

\newcommand{\mH}{\mathcal{H}}

\newcommand{\mL}{\mathcal{L}}

\begin{document}
%%%%%%%%%%%%%%%%%%%%%%%%%%%%%%%%%%
%%%%%%%%%%%%%%%%%%%%%%%%%%%%%%%%%%
\section{Introduction} Type IIB String theory on $AdS_5\times S^5$
has been conjectured to be dual to N=4 super Yang-Mills theory in
4-dimensions \cite{Maldacena:1997re}. The conjectured duality has
passed through various nontrivial tests in the past by analyzing
the spectrum of quantum string states on $AdS_5 \times S^5$
background and the spectrum of the anomalous dimensions of the N=4
gauge theory operators in the planar limit. Especially in the
semiclassical approximation the theory become {\it integrable} on
the both sides of the duality \footnote{For some works considering
integrability of sigma model on $AdS_5\times S^5$, see
\cite{Kluson:2007gp,
Dorey:2006mx,Gromov:2006dh,Alday:2005ww,Das:2005hp,Alday:2005jm,
Arutyunov:2005nk,Frolov:2005dj,Chen:2005uj,Alday:2005gi,Das:2004hy,
Arutyunov:2004yx,Beisert:2004ag,Kazakov:2004nh,Berkovits:2004jw,
Hatsuda:2004it,Kazakov:2004qf,Alday:2003zb,Vallilo:2003nx}. For
reviews of this problems from different point of views, see
\cite{Minahan:2006sk,Beisert:2004ry,Plefka:2005bk,
Tseytlin:2004xa,Zarembo:2004hp,Swanson:2005wz,Tseytlin:2003ii}.}.
 Though finding out the full
spectrum of string theory in AdS background is still an open
unsolved problem, it has been observed that in certain sectors of
the theory it is more tractable. One of them is the large angular
momentum sector of the theory \cite{Berenstein:2002jq}. In this
region, one can use the semiclassical approximations to find the
string spectrum as well \cite{Gubser:2002tv}. The correspondence
then demands to find out a particular type of operators that are
the long trace operators in the gauge theory side. This fact was
resolved by an amazing paper \cite{Minahan:2002ve} by relating the
Hamiltonian of a Heisenberg's spin chain system with that of the
dilatation operator in N=4 Supersymmetric Yang-Mills theory. So it
seems that there is a close interplay between string theory, gauge
theory and the spin-chain system, which makes the field of
research more fascinating.

One of the interesting topic that has drawn lot of attention
recently are the low lying spin chain system which correspond to
the magnon excitation. Hofman and Maldacena have proposed
\cite{Hofman:2006xt} not too long back the correspondence of these
magnon states with that of a specific configuration of
semiclassical string states on $R\times S^2$ \cite{Gubser:2002tv}.
In particular the giant magnon solution of correspond to operators
where one of the $SO(6)$ charge, $J$, is taken to infinity,
keeping the $E-J$ fixed\footnote{The Hofman-Maldacena limit:
$J\rightarrow \infty, \lambda = {\rm fixed}, p= {\rm fixed}, E-J =
{\rm fixed}$}. These excitations satisfy a dispersion relation of
the type (in the large ' t Hooft limit $(\lambda)$)
\begin{equation} E-J = \frac{\sqrt{\lambda}}{\pi}
\left|{\rm sin}\frac{p}{2}\right| \ ,
\end{equation}
where $p$ is the magnon momentum. Hence after lot of work has been
devoted to study and generalize to various other magnon states
with two and three non vanishing momentum and so on, see for
example \cite{Dorey:2006dq,Minahan:2006bd,Chen:2006ge,
Chu:2006ae,Spradlin:2006wk,Bobev:2006fg,Ryang:2006yq,Huang:2006vz}.
The basic idea in all these developments was to show what does
this correspond to in the gauge theory side of the duality
conjecture.

In the present paper we generalize the discussion of giant magnon
in the NS5-brane background. In string theory NS5-brane solution
is interesting because in the near horizon limit the theory on the
worldvolume correspond to a nonlocal field theory, namely the
little string theory (LST)
\footnote{For review, see \cite{Aharony:1999ks,Kutasov:2001uf}.}.
 Though the LST has been a misnomer
since long time, it seems a good exercise to analyze the solution
in various limits and find out the corresponding field theory. The
near horizon NS5-brane world sheet theory is exactly solvable
\cite{Callan:1992rs}, so from the bulk view point it seems that
the theory is integrable. A little is known about the boundary
theory, hence from that prospective it is rather hard to make
definite statements about the exact nature of the theory.

Motivated by the recent surge of interest in finding out the
magnon  solutions in anti-de sitter space, we study the NS5-brane
worldvolume theory in near horizon limit. The near horizon limit,
in a particular scaling yields a background with a linear dilaton.
In order to study the magnon solution, one has to take proper care
of the dilaton. We include the dilaton in the Polyakov action,
which modifies obviously the worldsheet stress tensor. We show
that even in the presence of a linear dilaton we get a giant
magnon solution, and the exact dispersion relation for the magnon
discussed in the literature.

Rest of the paper is organized as follows. In the section-2, we
write down the Polyakov string in the near horizon limit of the NS
5-brane. We find out the magnon dispersion relation in the large
$J$ sector. We get the dispersion relation in terms of total
energy $E$, momentum along an angular direction $(\phi_2)$ of the
sphere in the transverse direction of NS 5-brane $(P_{\phi_2})$,
momentum along the longitudinal directions of NS5-branes, $y^i$,
and another conserved charge $D$. The conserved charge $D$ is
related to the radial motion in the decoupling limit of
$NS5$-brane background that, with appropriate choice of
coordinates, is free and reflects the fact that the configuration
of NS5-brane and fundamental string is BPS state.

 In section-3, we try to study this situation further.
We develop a Hamiltonian formalism for string in the near horizon
region of $N$ NS5-branes with a general world-sheet metric.
However due to the presence of the nontrivial coupling of the
string to the dilaton, we take $\sqrt{\lambda}\rightarrow \infty$
to restrict to the study of the classical dynamics of string. Then
in the first approximation we can neglect the coupling of the
string to the dilaton in uniform gauge, and find the same magnon
dispersion relation of section-2. Finally in section-4, we present
our conclusion and outlook.
%%%%%%%%%%%%%%%%%%%%%%%%%%%%%%%%%%%%%%%%%%%%%%%
\section{Polyakov string in the near horizon limit of NS5-brane background}
% Let us consider the Nambu-Gotto action
% for the string in the general background
% \begin{equation}
% S=-\tau_{NS}
% \int_{-r}^r d\tau\sigma
% \sqrt{-\det\bA} \ ,
% \bA=\partial_\mu X^M\partial_\nu X^N
% G_{MN} \ .
% \end{equation}
In this section we study the Polyakov string in the near horizon
limit of the NS 5-brane background. The classical solution of $N$
NS5-brane is given by
\begin{eqnarray}
ds^2 &=& -d{\tilde t}^2 + dy^2_5 + H(r)\left(dr^2 + r^2
d\Omega^2_3\right) \cr & \cr e^{2(\Phi-\Phi_0)} &=& H(r), \>\>\>\>
b = 2 N \sin^2 \theta \cos \phi_1 d\theta \wedge d\phi_2, \>\>\> H
(r) = 1+ \frac{Nl^2_s}{r^2} \ ,
\end{eqnarray}
where $y^i \ , i=6,\dots,9$ label world-volume directions of
$NS5$-branes, $H(r)$ is the harmonic function in the transverse
space of the NS5-brane. Finally, $l_s$ is string length that
is related to Regge slope as $l_s^2=\frac{1}{\alpha'}$.

In the near horizon limit, $r \rightarrow 0$, one can ignore the 1
in $H(r)$, and solution would look like (defining ${\tilde t} =
\sqrt{N}l_s t$)
\begin{equation}\label{NS5bac}
ds^2=Nl_s^2(-dt^2+d\theta^2+ \sin^2\theta
d\phi^2_1+\sin^2\theta\sin^2\phi_1
d\phi^2_2+\frac{dr^2}{r^2})+dy_5^2 \ ,
\end{equation}
where the metric on $S^3$ is written as
\begin{equation}
d\Omega^2_3=d\theta^2+\sin^2\theta d\phi^2_1+\sin^2\theta
\sin^2\phi_1  d\phi^2_2 \ .
\end{equation}
At the same time
\begin{equation}\label{NS5bacd}
e^{2\Phi}=\frac{Nl_s}{r^2} \ \ , \quad b_{\theta \phi_2} = 2N
\sin^2 \theta \cos \phi_1 \ .
\nonumber \\
\end{equation}
We are interested in the geometry (\ref{NS5bac}), for finding out
the giant magnon solution that is analogue of the giant magnon
solution in $AdS_5 \times S^5$ found in \cite{Hofman:2006xt}. Our
starting point is the Polyakov form of the string action in the
background (\ref{NS5bac})
\begin{eqnarray}\label{actPol}
S&=&-\frac{\sqrt{\lambda}}{4\pi}
\int_{-\pi}^\pi d\sigma d\tau
[\sqrt{-\gamma}\gamma^{\alpha\beta}
g_{MN}\partial_\alpha x^M\partial_\beta x^N
-e^{\alpha\beta}
\partial_\alpha x^M\partial_\beta x^N b_{MN}]+
\nonumber \\
&+&\frac{1}{4\pi}\int_{-\pi}^\pi
d\sigma d\tau \sqrt{-\gamma}
R\Phi \ , \nonumber \\
\end{eqnarray}
where the pre-factor $\sqrt{\lambda}$ is equal to
\begin{equation}
\sqrt{\lambda}=N \ ,
\end{equation}
and where $\gamma^{\alpha\beta}$ is
world-sheet metric and $R$ is its
Ricci scalar. Further, $e^{\alpha\beta}$
is defined as $e^{01}=-e^{10}=1$. Finally,
 the modes
$x^M, M=0,\dots,9$ parameterize the embedding of the string in the
background (\ref{NS5bac}).
% The equation of motions for
% $x^M$ that follow from (\ref{actPol})
% take the form
% \begin{eqnarray}
% -g_{LK}[
% \partial_\alpha\left(\sqrt{-\gamma}
% \gamma^{\alpha\beta}\partial_\beta X^K\right)
% +\sqrt{-\gamma}\gamma^{\alpha\beta}
% \Gamma^K_{MN}\partial_\alpha X^M
% \partial_\beta X^N]=\nonumber \\
% =\frac{1}{2}H_{LMN}\epsilon^{\alpha\beta}
% \partial_\alpha X^N\partial_\beta X^N \ ,
% H=dB
% \nonumber \\
% \end{eqnarray}
% where
% \begin{equation}
% \Gamma_{MNP}=
% \frac{1}{2}(
% g_{MN,P}+g_{MP,N}-g_{NP,M})
% \end{equation}
% \begin{equation}
% \partial_\mu [G_{MN}
% \partial_\nu X^N\bAi^{\nu\mu}
% \sqrt{-\det\bA}]-
% \frac{1}{2}\partial_M g_{KL}\partial_\mu
% X^K\partial_\nu X^L \bAi^{\nu\mu}
% \sqrt{-\det\bA}=0
% \end{equation}
% Now we consider this string in the near
% horizon limit of NS5-brane background
The variation of the action
(\ref{actPol})
with respect to $x^M$ implies
following equations of motion
\begin{eqnarray}
-\frac{\sqrt{\lambda}}
{4\pi}\sqrt{-\gamma}\gamma^{\alpha\beta}
\partial_K
g_{MN}\partial_\alpha x^M\partial_\beta x^N
+\frac{\sqrt{\lambda}}{2\pi}\partial_\alpha[
\sqrt{-\gamma}\gamma^{\alpha\beta}
g_{KM}\partial_\beta x^M]-
\nonumber \\
-\frac{\sqrt{\lambda}}{2\pi}
\partial_\alpha[
\epsilon^{\alpha\beta}
\partial_\beta x^M b_{KM}]
+\frac{\sqrt{\lambda}}{4\pi}
\epsilon^{\alpha\beta}
\partial_\alpha x^M\partial_\beta x^N
\partial_K b_{MN}
+\frac{1}{4\pi}\partial_K
\Phi \sqrt{-\gamma}R=0 \ .
\nonumber \\
\end{eqnarray}
Further, the variation of the action with
respect to the metric implies the constraints
\cite{Rudd:1994ss}
% and the constraints
% \begin{equation}
% (\gamma^{\alpha\beta}
% \gamma^{\gamma\delta}
% -2\gamma^{\alpha \gamma}
% \gamma^{\beta\delta})
% \partial_\gamma X^M\partial_\delta
% X^N g_{MN}=0
% \end{equation}
\begin{eqnarray}\label{gravcons}
-\frac{4\pi}{\sqrt{-\gamma}}
\frac{\delta S}{\delta \gamma^{\alpha
\beta}}&=&
\sqrt{\lambda}
g_{MN}\partial_\alpha x^M\partial_\beta x^N-R_{\alpha
\beta}+
\nonumber \\
&+&(\nabla_\alpha \nabla_\beta x^M)
\partial_M \Phi+(\partial_\alpha x^M\partial_\beta
x^N)\partial_M\partial_N\Phi
-\nonumber \\
&-&\frac{1}{2}
\gamma_{\alpha\beta}
\left(\sqrt{\lambda}\gamma^{\gamma\delta}
\partial_\gamma x^M\partial_\delta
x^N g_{MN}-R\Phi+2
\nabla^\alpha \nabla_\alpha \Phi\right) \ .
\nonumber \\
\end{eqnarray}
% using
% \begin{eqnarray}
% \delta (\sqrt{-\gamma})=
% -\frac{1}{2}\gamma_{\alpha\beta}
% \delta \gamma^{\alpha\beta}\sqrt{-\gamma} \ ,
% \nonumber \\
% \delta R=
% -\frac{1}{2}R\gamma^{\alpha\beta}\delta
% \gamma_{\alpha\beta}-\frac{1}{2}
% \nabla_\alpha \nabla^\alpha (\gamma^{
% \beta\gamma}\delta \gamma_{\beta\gamma})
% +\frac{1}{2}\nabla^\alpha \nabla^\beta
% \delta g_{\alpha\beta} \ ,
% \nonumber \\
% \nabla_\alpha V_\beta=
% \partial_\alpha-\Gamma_{\alpha\beta}^\gamma V_\gamma \ ,
% \quad
% \nabla_\alpha V^\beta=
% \partial_\alpha V^\beta+\Gamma^\beta_{\alpha
% \gamma}V^\gamma \ , \nonumber \\
% \end{eqnarray}
To proceed further it is convenient
to introduce the variable $\rho$
that is related to $r$ as
\begin{equation}\label{rhor}
\rho=\ln \left(\frac{r}{\sqrt{Nl_s^2}}\right) \ .
\end{equation}
% so that
% \begin{equation}
% g_{rr}\partial_\alpha r\partial_\beta r=
% \partial_\alpha \rho\partial_\beta \rho \ ,
% \quad \Phi=-\rho \ ,
% \quad \sqrt{-\gamma}R\Phi=-\sqrt{-\gamma}R\rho \ .
% \end{equation}
Then in the  conformal gauge $
\gamma_{\alpha\beta}=\eta_{\alpha\beta}$ where
$\eta_{\alpha\beta}=
\mathrm{diag}(-1,1)$ and using the variable
(\ref{rhor}) the constraints
 (\ref{gravcons}) take  simpler form
\begin{eqnarray}\label{Tcon}
T_{\sigma\sigma}&=&-4\pi
\frac{\delta S}{\delta \gamma^{\sigma
\sigma}}=
\frac{\sqrt{\lambda}}{2}
(g_{MN}\partial_\sigma x^M\partial_\sigma x^N
+g_{MN}\partial_\tau x^M\partial_\tau x^N)-
\partial_\tau^2 \rho \ ,
\nonumber \\
T_{\tau\tau}&=&-4\pi
\frac{\delta S}{\delta \gamma^{\tau
\tau}}=
\frac{\sqrt{\lambda}}{2}
(g_{MN}\partial_\sigma x^M\partial_\sigma x^N
+g_{MN}\partial_\tau x^M\partial_\tau x^N)-
\partial_\sigma^2 \rho \ ,
\nonumber \\
T_{\tau\sigma}&=&-4\pi
\frac{\delta S}{\delta \gamma^{\sigma\tau
}}=
\sqrt{\lambda}
g_{MN}\partial_\sigma x^M\partial_\tau x^N
-\partial_\sigma\partial_\tau \rho \ .
\nonumber \\
\end{eqnarray}
Looking at the form of the background (\ref{NS5bac}) and
(\ref{NS5bacd}) we observe that the action (\ref{actPol})
 is invariant under
following transformations of fields
\begin{eqnarray}
t'(\tau,\sigma)&=&t(\sigma,\tau)+\epsilon_t \ ,
\nonumber \\
\phi'_{2}(\tau,\sigma)&=&
\phi_{2}(\tau,\sigma)+\epsilon_{\phi_2} \ ,
\nonumber \\
\end{eqnarray}
where $\epsilon_t,\epsilon_{\phi_2}$ are constants. Then it is a
simple task to determine corresponding conserved charges
\begin{eqnarray}\label{CGcon}
P_t&=&\frac{\sqrt{\lambda}}{2\pi}
\int_{-\pi}^\pi d\sigma
\sqrt{-\gamma}\gamma^{\tau\alpha}
g_{tt}\partial_\alpha t \ , \nonumber \\
P_{\phi_2}&=&
\frac{\sqrt{\lambda}}{2\pi}
\int_{-\pi}^\pi d\sigma
[\sqrt{-\gamma}\gamma^{\tau\alpha}
g_{\phi_2\phi_2}\partial_\alpha \phi_2
-\partial_\alpha\theta b_{\theta\phi_2}] \ .
\nonumber \\
\end{eqnarray}
Note that $P_t$ is related to the energy as
$P_t=-E$.

Moreover, the string action in the conformal
gauge in the near horizon
limit of $NS5$-brane background is also invariant
under following transformation
\begin{equation}
r'(\tau,\sigma)=\Lambda r(\tau,\sigma)
\end{equation}
for constant $\Lambda$. Alternatively, using the variable $\rho$
this symmetry is an ordinary translation symmetry
\begin{equation}
\rho'(\tau,\sigma)=
\rho(\tau,\sigma)+\epsilon_\rho \
\end{equation}
for constant $\epsilon_\rho$.
This invariance leads to the
existence of the third conserved charge
\begin{equation}\label{Dcharge}
D=\frac{\sqrt{\lambda}}{2\pi} \int_{-\pi}^{\pi}d\sigma
\sqrt{-\eta}\eta^{\tau\alpha} \frac{1}{r}\partial_\alpha r
\end{equation}
or
\begin{equation}\label{Dchargerho}
D=\frac{\sqrt{\lambda}}{2\pi} \int_{-\pi}^\pi d\sigma
\sqrt{-\eta}\eta^{\tau\alpha}
\partial_\alpha \rho \ .
\end{equation}
Finally, the invariance of the metric
with respect of the translation in $y^i$-directions
implies an existence of five conserved charges
\begin{equation}\label{Pyi}
P_{y_i}=
\frac{\sqrt{\lambda}}{2\pi}
\int_{-\pi}^\pi d\sigma
\sqrt{-\eta}\eta^{\tau\alpha}
\partial_\alpha y^i \ .
\end{equation}
Now we proceed to find the solution of the equations of motion
given above which could be interpreted as an giant magnon. We
closely follow very nice analysis presented in
\cite{Arutyunov:2006gs}. (See also \cite{Astolfi:2007uz}).

Let us now consider following ansatz for obtaining the giant
magnon solution
\begin{eqnarray}\label{ans}
t=-\frac{E}{\sqrt{\lambda}} \tau \ , \quad
\theta=\theta(\sigma,\tau), \quad  \rho=\rho(\tau)\ , \nonumber \\
%  r=r(\tau) \ ,
\phi_2=\phi_2(\sigma,\tau) \ , \quad \phi_1=const. ,
 \quad y^i=v^i\tau \ . \nonumber \\
\end{eqnarray}
% We will solve this ansatz in the conformal
% gauge where $\gamma^{\tau\tau}=-1 \ ,
% \gamma^{\tau\sigma}=1 \ , \gamma^{\tau\sigma}=0$.
In conformal gauge the equation of motion for $\rho$ takes the
form
\begin{equation}\label{Req}
\partial_\alpha
[\sqrt{-\eta}
\eta^{\alpha\beta}\partial_
\beta \rho]=0
% +\partial_R
% G_{rr}\dot{R}^2
% -2\partial_\tau[
% G_{rr}\dot{R}]=\Rightarrow
% \nonumber \\
% \ddot{R}-\frac{\dot{R}^2}{R}=\Rightarrow
% \nonumber \\
% \partial_\tau [\ln \dot{R}]=
% \partial_\tau \ln R \Rightarrow
% \nonumber \\
% \ln \dot{R}=\ln C+\ln R \Rightarrow
% R=R_0 e^{Ct} \nonumber \\
\end{equation}
with the general solution
\begin{equation}\label{rhosol}
\rho=C\tau+\rho_0 \ ,
\end{equation}
where  $C,\rho_0$ are constants.
 In order to assure that we are in the weak
coupling region we have to demand that $C>0$ and also
consider large $\rho_0$.

On the other hand
the equation of motion for $t$ takes the form
\begin{eqnarray}
\partial_\alpha[\sqrt{-\eta}\eta^{\alpha\beta}
g_{tt}\partial_\beta t]=0
\nonumber \\
\end{eqnarray}
that is clearly solved with the
ansatz (\ref{ans}). It can be also easily
shown that the equations of motion for $y^i$
are solved with the ansatz (\ref{ans}).
% The equation of motion for $\theta$ takes
% the form
% \begin{eqnarray}
% -\partial_\theta G_{\theta\theta}
% [\theta'^2-\dot{\theta}^2]-
% \partial_\theta G_{\phi_1\phi_1}
% (\partial_\sigma\phi_1)^2
% -\partial_\theta G_{\phi_2\phi_2}
% [\phi_2'^2-\dot{\phi}_2^2]+
% \nonumber \\
% +2[\theta''-\ddot{\theta}]
% +2\partial_\sigma [\dot{\phi}_2
% B_{\theta\phi_2}]-2\partial_\tau[
% \phi_2'B_{\theta\phi_2}]
% -2\theta'\dot{\phi_2}\partial_\theta B_{\theta\phi_2}
% +2\dot{\theta}\phi_2'\partial_\theta B_{\theta\phi_2}
% =0
% \Rightarrow
% \nonumber \\
% -2\sin\theta\cos\theta\sin^2\phi_1
% (\phi_2')^2+2
% \sin\theta\cos\theta \sin^2\phi_1\dot{\phi}_2^2
% +2[\theta''-\ddot{\theta}]=0 \ ,  \nonumber \\
% \end{eqnarray}
Further it is easy
to see that the
 equation of motion for
 $\phi_1$  has two constant
 solutions:
 \begin{equation}
\phi_1=0 \ , \quad
\phi_1=\frac{\pi}{2} \ .
\end{equation}
Solution with $\phi_1=0$ corresponds to the trivial pint-like
solution and we will not consider it further. We will restrict
ourselves to the case when $\phi_1 =\frac{\pi}{2}$.
 Note also that for $\phi_1=\frac{\pi}{2}$ the $b$ field
 vanishes.
% \begin{eqnarray}
% \partial_{\phi_1} G_{\phi_2\phi_2}
% [(\partial_\tau \phi_2)^2-
% (\partial_\sigma \phi_2)^2]=0
% \Rightarrow 2\sin \phi_1\cos \phi_1=0
% \nonumber \\
% \end{eqnarray}
% Finally the equation of motion for $\phi_2$
% takes the form
% \begin{eqnarray}
% 2\partial_\sigma[G_{\phi_2\phi_2}
% \phi'_2]
% -2\partial_\tau[G_{\phi_2\phi_2}\dot{\phi}_2]
% +\nonumber \\
% +2\partial_\tau[\theta' B_{\theta\phi_2}]-
% 2\partial_\sigma[\dot{\theta}B_{\theta\phi_2}]
% \Rightarrow
% \nonumber \\
% \sin^2\theta
% \sin^2\phi_1
% [\phi_2''-\ddot{\phi}_2]+
% 2\sin\theta\cos\theta
% \sin^2\phi_1
% [\theta'\phi_2'-\dot{\theta}\phi_2']
% =0 \ .
% \nonumber \\
% \end{eqnarray}

Instead to solve the equations of motions for $x$ directly it is
more convenient to solve the constraints (\ref{Tcon}) that can be
interpreted as the first integrals of the theory. Inserting
 the ansatz (\ref{ans}) to   the constraints
coming from the variation of
the metric (\ref{Tcon}) we obtain two
equations
\begin{eqnarray}\label{Tconas}
% T_{\sigma\sigma}&=&T_{\tau\tau}=
% \frac{1}{2}
% (\sqrt{\lambda}
% g_{MN}\partial_\sigma x^M\partial_\sigma x^N
% +g_{MN}\partial_\tau x^M\partial_\tau x^N)-
% \partial_\tau^2 \rho=
% \nonumber \\
% =\frac{1}{2}(\partial_\tau X^M\partial_\tau X^N g_{MN}+
% \partial_\sigma X^M\partial_\sigma X^N g_{MN})=0
% \Rightarrow
% \nonumber \\
& &-\frac{E^2}{\lambda}
+C^2+v_iv^i+
\sin^2\theta
(\dot{\phi_2})^2+
\dot{\theta}^2+\theta'^2+
\sin^2\theta\phi_2'^2=0 \ ,
\nonumber \\
& &
% g_{MN}\partial_\sigma x^M\partial_\tau x^N
% =0 \Rightarrow
% \nonumber \\
\dot{\phi}_2\phi'_2
\sin^2\theta+
\theta'\dot{\theta}=0 \ ,
\nonumber \\
\end{eqnarray}
where we have used
(\ref{rhosol})
and also the notation
$\partial_\sigma x=x',
\partial_\tau x=\dot{x}$.
% Let us discuss the solutions of these equations.
% For $\phi_1=0$ the last equation implies
% that $\theta'=0$ or $\dot{\theta}=0$.
% The equation of motion for $\phi_2$ is
% trivially satisfied while the equation
% of motion for $\theta$ implies
% \begin{equation}
% \theta'=\pm \dot{\theta}
% \end{equation}
% that thanks to the second Virasoro constraints
% implies
% \begin{equation}
% \theta'=\dot{\theta}=0
% \end{equation}
% In other words the solution with $\phi_1=0$ leads
% to the point-like solution and we will not discuss
% it further.
% Let us then restrict to the
% case $\phi_1=\frac{\pi}{2}$.
 Following \cite{Hofman:2006xt,Arutyunov:2006gs}
 we   search for a
solution with the boundary conditions
\begin{equation}
\theta(\pi,\tau)-\theta(-\pi,\tau)=0 \ , \quad \triangle \phi_2=
\phi_2(\pi,\tau)-\phi_2(-\pi,\tau)=p \ ,
\end{equation}
where $p$ is the momentum of the `single magnon' excitation. Since
the field $\phi_2$ does not satisfy periodic boundary conditions
this solution corresponds to the open string.

As the next step we introduce the light-cone
coordinate $\varphi$ through the formula
\begin{equation}
\phi_2=\varphi_2+\omega \tau
\end{equation}
and take the ansatz for $\varphi_2$ and
$\theta$ in the form
\begin{equation}
\theta=\theta(\sigma-v\omega \tau) \ ,
\quad \varphi_2=\varphi_2(\sigma-v\omega \tau) \ .
\end{equation}
With this ansatz eqn. (\ref{Tconas}) take the forms
\begin{eqnarray}\label{Virins}
& &(\omega-v\omega\varphi'_2)
\varphi'_2\sin^2\theta-
v\omega \theta'^2=0 \ ,
\nonumber \\
& &-\frac{E^2}
{\lambda}+C^2+v_iv^i+\sin^2\theta(\omega-v\omega \varphi'_2)^2+
(1+v^2\omega^2)\theta'^2+\sin^2\theta\varphi_2'^2=0 \ ,
\nonumber \\
\end{eqnarray}
where now $\theta',\varphi_2'$ means derivative with respect of
arguments of $\theta,\varphi_2$ respectively. If we now combine
these equations we obtain
\begin{eqnarray}\label{sol}
% \theta'^2=
% \frac{(1-v\varphi_2')\varphi_2'\sin^2\theta}{v} \ ,
% \nonumber \\
% -\frac{E^2}{\lambda}
% +C^2+\sin^2\theta \omega^2(1-v\varphi'_2)^2+
% \sin^2\theta \varphi_2'^2+
% \frac{(1+v^2\omega^2)}{v}(1-v\varphi_2')
% \varphi_2'\sin^2\theta=0 \Rightarrow
% \nonumber \\
% -\frac{E^2}{\lambda}
% +C^2+[\omega^2+\frac{(1-v^2\omega^2)}{v}\varphi_2']
% \sin^2\theta=0 \Rightarrow \nonumber \\
\varphi_2'&=&\frac{v(\frac{E^2}
{\lambda}-C^2-v_iv^i-\omega^2\sin^2\theta)}
{(1-v^2\omega^2 )\sin^2\theta}  \ , \nonumber \\
\theta'^2&=&
% \frac{(\frac{E^2}{\lambda}-C^2-\omega^2\sin^2\theta)}
% {(1-v^2\omega^2 )}
% -[\frac{v(\frac{E^2}{\lambda}
% -C^2-\omega^2\sin^2\theta)}
% {(1-v^2\omega^2 )\sin^2\theta}]^2
% \sin^2\theta=
%  \nonumber \\
%  =\frac{\sin^2\theta(\frac{E^2}{\lambda}
% -C^2)(1+v^2\omega^2)
%  -\omega^2\sin^4\theta-v^2(\frac{E^2}{\lambda}
% -C^2)^2}
%  {(1-v^2\omega^2)^2\sin^2\theta}= \nonumber \\
\omega^2\frac{(\sin^2\theta_{max}-
\sin^2\theta)(
\sin^2\theta-\sin^2\theta_{min})}
{(1-v^2\omega^2)^2\sin^2\theta}
\ ,   \nonumber \\
\end{eqnarray}
where
% Now the condition that $\theta^2$
% is positive implies following bound for
% $\theta$
\begin{eqnarray}
% -\sin^2\theta\frac{1}{\omega^2}
% (\frac{E^2}{\lambda}-C^2)(1+v^2\omega^2)+
% \sin^4\theta+\frac{v^2}{\omega^2}(\frac{E^2}{\lambda}
% -C^2)^2<0
%  \Rightarrow \nonumber \\
% \sin^2\theta_{min}<\sin^2\theta <
% \sin^2\theta_{max} \ ,
\sin^2\theta_{max}=
 \frac{\frac{E^2}{\lambda}-C^2-v_iv^i}{\omega^2} \ , \quad
 \sin^2\theta_{min}=\left(\frac{E^2}{\lambda}
-C^2-v_iv^i\right)v^2 \ .
 \nonumber \\
\end{eqnarray}
We see, in agreement with
\cite{Arutyunov:2006gs},  that for
this solution the derivative $\theta'$ is
finite everywhere and vanishes both
at $\sin^2\theta_{max}$ and
$\sin^2\theta_{min}$.
On the other hand using the arguments
given there we can ague that the
target space shape of the solution
is determined by $\frac{d\theta}{d\varphi_2}$
that vanishes at $\theta=\theta_{max}$
and diverges at $\theta=\theta_{min}$.

We can also express the velocity $v$
in terms of $p$ by using the boundary condition
for $\phi_2$  since
\begin{equation}
\triangle \phi_2=
\int_{-\pi}^\pi
d\sigma \partial_\sigma \phi=
\int_{-\pi}^\pi
d\sigma \varphi'_2=
\int_{\theta_{min}}^{\theta_{max}}
d\theta \frac{\varphi'_2}{|\theta'|} \ .
 \end{equation}
% Let us check the equation of motion
% for this ansatz. The equation of motion
% for $\phi_2$ implies
% \begin{eqnarray}
% [g_{\phi_2\phi_2}\varphi_2']'+
% \omega v [g_{\phi_2\phi_2}(\omega-v\omega\varphi_2')]'=0
% \Rightarrow
% \nonumber \\
% \sin^2\theta[\varphi_2''-v^2\omega^2\varphi_2'']+
% 2\sin\theta\cos\theta
% [\theta'\varphi_2'-v\omega \theta'\varphi_2'
% +v\omega^2]=0
% \nonumber \\
% \varphi_2'=\frac{K}{\sin^2\theta
% (1-v^2\omega^2)}-\frac{\omega^2 v}{1-v^2\omega^2}
% \nonumber \\
% \end{eqnarray}
% that coincides with the equation for $\varphi_2$
% that were determined from the Virasoro constraints
% with appropriate identification of constant.
% On the other hand the equation of motion for $\theta$
% takes the form
% \begin{eqnarray}
% -\partial_\theta g_{\phi_2\phi_2}
% [\varphi_2'^2-\omega^2(1-v\varphi_2')^2]
% +2(1-v^2\omega^2)\theta''=0 \ \quad
% \times \theta' \Rightarrow
% \nonumber \\
% -g_{\phi_2\phi_2}'
% [\varphi_2'^2-\omega^2(1-v\varphi_2')^2]+
% (1-v^2\omega^2)(\theta'^2)'=0
% \nonumber \\
% \end{eqnarray}
% {\bf Check it!!!}
As the next step we insert
(\ref{sol}) into
 the conserved charges given in
(\ref{CGcon}) and we obtain
\begin{eqnarray}
P_t&=&\frac{\sqrt{\lambda}}{2\pi} \int_{-\pi}^\pi
d\sigma[\sqrt{-\eta}\eta^{\tau\alpha} g_{tt}\partial_\alpha t]=-E
\ ,
\nonumber \\
P_{\phi_2}&=&\frac{\sqrt{\lambda}}{2\pi}
\int_{-\pi}^\pi
d\sigma[\sqrt{-\eta}
\eta^{\tau\alpha}g_{\phi_2\phi_2}
\partial_\alpha \phi_2-\partial_\sigma
\theta b_{\theta\phi_2}]=\nonumber \\
% =2\frac{\sqrt{\lambda}}{2\pi}
% \int_{\theta_{min}}^{\theta_{max}}d\theta
% \frac{\omega}{|\theta'|}
% \sin^2\theta
% (v\varphi'_2-1))
% -2\frac{\sqrt{\lambda}}{2\pi}
% \int_{\theta_{min}}^{\theta_{max}}
% d\theta B_{\theta \phi_2}
% =
% \nonumber \\
% =|1-z^2=\sin^2\theta|=
% 2\frac{\sqrt{\lambda}}{2\pi}
% \int_{z_{min}}^{z_{max}}dz
% \frac{\omega (1-z^2)}{|z'|}(v\varphi'_2-1)=
% \nonumber \\
% -\frac{\sqrt{\lambda}}{\pi}
% \int_{\theta_{min}}^{\theta_{max}}
% d\theta(1-\cos 2\theta)=
% \nonumber \\
&=&\frac{\sqrt{\lambda}}{\pi}
\int_{z_{min}}^{z_{max}}dz
\frac{\omega (1-z^2)}{|z'|}(v\varphi'_2-1)
 \ , \nonumber \\
% -
% \frac{\sqrt{\lambda}}{\pi}
% (\theta_{max}-\theta_{min})+
% \frac{\sqrt{\lambda}}{2\pi}
% [\sin 2\theta_{max}-\sin 2\theta_{min}]
\end{eqnarray}
where we  introduced the coordinate
$z$ defined as
\begin{equation}
1-z^2=\sin^2\theta \ .
\end{equation}
Finally we insert (\ref{ans})
into (\ref{Dcharge}) and (\ref{Pyi})
and we obtain
 \begin{eqnarray}\label{Dins}
D=\frac{\sqrt{\lambda}}{2\pi} \int_{-\pi}^{\pi}
d\sigma\sqrt{-\eta}\eta^{\tau\alpha} \frac{1}{r}\partial_\alpha r=
\sqrt{\lambda}C \ , \nonumber \\
 P_{y_i}=\frac{\sqrt{\lambda}}{2\pi}
\int_{-\pi}^{\pi} d\sigma\sqrt{-\eta}\eta^{\tau\alpha}
\partial_\alpha y^i=\sqrt{\lambda}v^i \ .
\nonumber \\
\end{eqnarray}
The next goal is to explicitly integrate these expressions.
Following the analysis given in  \cite{Arutyunov:2006gs} we easily
get
\begin{eqnarray}\label{ptriangle}
p&=&\triangle \phi_2=
% \int_{-\pi}^\pi
% d\sigma \partial_\sigma \phi=
% \int_{-\pi}^\pi
% d\sigma \varphi'_2=
2\int_{\theta_{min}}^{\theta_{max}}
d\theta \frac{\varphi'_2}{|\theta'|}=
2\int_{z_{min}}^{z_{max}}
\frac{dz}{|z'|}\varphi'_2=\nonumber \\
&=& 2v\omega\int_{z_{min}}^{z_{max}}
dz \frac{z^2-z_{min}^2}{
(1-z^2)\sqrt{z_{max}^2-z^2}\sqrt{z^2-z_{min}^2}}=
\nonumber \\
% =-2\frac{v\omega}{z_{max}}K(\eta)+
% 2\frac{v\omega(1-z_{min}^2)}{z_{max}
% (1-z^2_{max})}\Pi
% \left(\frac{z^2_{max}-z^2_{min}}{z^2_{max}-1},
% \eta\right)=
% \nonumber \\
&=&-2\frac{v\omega}{\sqrt{
1-\left(\frac{E^2}{\lambda}-C^2-v_iv^i
\right)v^2}}K(\eta)+\frac{2
\Pi\left(\frac{v^2\left(\frac{E^2}{\lambda}
-C^2-v_iv^i\right)-1}{
\left(\frac{E^2}{\lambda}-C^2-v_iv^i
\right)v^2}\eta,\eta\right)}{
\omega v \sqrt{1-\left(\frac{E^2}{\lambda}-C^2
-v_iv^i
\right)v^2}} \ ,
\nonumber \\
\end{eqnarray}
% using the fact that
% \begin{eqnarray}
% \frac{d\theta}{|\theta'|}=
% -\frac{z dz}{\sin\theta\cos\theta
% |\frac{d\theta}{dz}||z'|}=
% \frac{dz}{|z'|} \ , \nonumber \\
% z'^2=(1-z^2)\theta'^2=(1-z^2)
% \omega^2\frac{(z^2_{max}-z^2)(z^2_{min}-z^2)}
% {(1-v^2\omega^2)^2(1-z^2)}=
% \omega^2\frac{(z^2_{max}-z^2)(z^2_{min}-z^2)}
% {(1-v^2\omega^2)^2} \ ,
%   \nonumber \\
% \varphi_2'=\frac{v\omega^2(z^2-z_{min}^2)}
% {(1-v^2\omega^2)
% (1-z^2)} \ , \nonumber \\
% \end{eqnarray}
where
\begin{eqnarray}
z_{max}^2&=&1-\sin^2\theta_{max}=
1-\left(\frac{E^2}{\lambda}-C^2-v_iv^i
\right)v^2 \ ,
\nonumber \\
z_{min}^2&=&1-\sin^2\theta_{min}=
1-\frac{\frac{E^2}{\lambda}-C^2-v_iv^i}
{\omega^2} \ , \nonumber \\
\eta&=& 1-\frac{z_{min}^2}{z_{max}^2}
\ . \nonumber \\
\end{eqnarray}
In the same way we can
integrate  $P_{\phi_2}$
with the result
\begin{eqnarray}\label{Pphi2}
P_{\phi_2}&=&J=-2\frac{\sqrt{\lambda}}{2\pi}
\int_{z_{min}}^{z_{max}}dz
\frac{\omega (1-z^2)}{|z'|}(1-v\varphi'_2)=
\nonumber \\
% +\frac{\sqrt{\lambda}}{\pi}
% (\theta_{max}-\theta_{min})+
% \frac{\sqrt{\lambda}}{2\pi}
% [\sin 2\theta_{max}-\sin 2\theta_{min}]
% =-\frac{\omega}{|\omega|}
% \frac{\sqrt{\lambda}}{\pi}
% \int_{z_{min}}^{z_{max}}
% \frac{dz}{\sqrt{z^2-z_{min}^2}\sqrt{
% z_{max}^2-z^2}}(1-z^2+\omega^2v^2(z^2_{min}-1)
% +\nonumber \\
% +\frac{\sqrt{\lambda}}{\pi}
% (\theta_{max}-\theta_{min})+
% \frac{\sqrt{\lambda}}{2\pi}
% [\sin 2\theta_{max}-\sin 2\theta_{min}]
% =\nonumber \\
% =-\frac{\omega}{|\omega|}\frac{\sqrt{\lambda}}{\pi}
% \frac{1}{z_{max}}K(\eta)+\frac{\sqrt{\lambda}}{\pi}
% z_{max}E(\eta)
% -\frac{\sqrt{\lambda}}{\pi}
% \frac{\omega^2 v^2(z^2_{min}-1)}{z_{max}}
% K(\eta)
% +\nonumber \\
% +\frac{\sqrt{\lambda}}{\pi}
% (\theta_{max}-\theta_{min})+
% \frac{\sqrt{\lambda}}{2\pi}
% [\sin 2\theta_{max}-\sin 2\theta_{min}]
&=&-\frac{\omega}{|\omega|}\frac{\sqrt{\lambda}}{\pi}
\sqrt{1-v^2\left(\frac{E^2}{\lambda}-C^2-v_iv^i
\right)}(K(\eta)-E(\eta)) \ .
% +\nonumber \\
% +\frac{\sqrt{\lambda}}{\pi}
% (\theta_{max}-\theta_{min})+
% \frac{\sqrt{\lambda}}{2\pi}
% [\sin 2\theta_{max}-\sin 2\theta_{min}]
\nonumber \\
\end{eqnarray}
Note that  the functions $E(\eta), K(\eta),$ and $\Pi$
are defined  by formulas
(We follow the notation given
in \cite{Arutyunov:2006gs}.)
\begin{eqnarray}
I_1&=&\int_{z_{min}}^{z_{max}}
dz\frac{1}{\sqrt{z^2-z_{min}^2}
\sqrt{z_{max}^2-z^2}}=\frac{1}{z_{max}}
K(\eta) \ , \nonumber \\
I_2&=& \int_{z_{min}}^{z_{max}}
dz\frac{z^2}{\sqrt{z^2-z^2_{min}}
\sqrt{z^2_{max}-z^2}}=z_{max}E(\eta) \ ,
\nonumber \\
I_3&=& \int_{z_{min}}^{z_{max}}
dz\frac{1}{(1-z^2)
\sqrt{z^2-z^2_{min}}\sqrt{z^2_{max}-z^2}}
=\frac{1}{z_{max}(1-z_{max}^2)}
\Pi\left(\frac{z_{max}^2-z^2_{min}}{
z_{max}^2-1},\eta\right) \ .
\nonumber \\
\end{eqnarray}
The relation (\ref{Pphi2}) allows us to express the modulus $\eta$
in terms of $J\equiv P_{\phi_2}, E, v, C$ and $v_i$.
 On the other hand it is clear that
there is no analytic expression for the dispersion relation.
However we would like to analyze these equations for large values
of charge $J$.

% In order to consider limit $J\rightarrow \infty$
% we introduce the variable $\epsilon=1-\eta$.
% Obviously $\epsilon\rightarrow 0$ for
% $\eta\rightarrow 1$. Note that this limit
% corresponds $z_{min}\rightarrow 0$.
%%%%%%%%%%%%%%%%%%%%%%%%%%%%%%%%%%%%%%%%%%%%%%%%%%%%%%
\subsection{Infinite $J$ Giant Magnon} To find more illuminating
result we restrict to the case of an infinite $J$ giant magnon.
This situation occurs for $z_{min}\rightarrow 0$ that corresponds
to
\begin{equation}
\omega^2=\frac{E^2}{\lambda}-C^2-v_iv^i \ .
\end{equation}
In fact, it is easy to see that
for $z_{min}\rightarrow 0$ the
charge
(\ref{Pphi2}) really diverges.
%\begin{equation}
%P_{\phi_2}=
%-\frac{\omega}{|\omega|}\frac{\sqrt{\lambda}}{\pi}
%\int_0^{z_{max}}dz\frac{z^2_{max}-z^2}{
%z\sqrt{z^2_{max}-z^2}}=
%\frac{\omega}{|\omega|}\frac{\sqrt{\lambda}}{\pi}
%\int_0^{z_{max}}dz\frac{z^2_{max}-z^2}{
%z\sqrt{z^2_{max}-z^2}} \
%\end{equation}
%that is divergent.
We  also  choose $\omega=-\sqrt{\frac{E^2}{\lambda}- C^2-v_iv^i}$
in order to have positive, large $J$.

Our goal now is to find the dispersion relation between $E,J$ and
possibly another charges that is an analogue of the relation found
in \cite{Hofman:2006xt}. It turns out that the relevant dispersion
relation takes the form
\begin{eqnarray}\label{di1}
& &\sqrt{E^2-D^2-(P_{y_i})^2}-J=
\frac{\sqrt{\lambda}}{\pi}
\int_0^{z_{max}}dz\frac{ z}{
\sqrt{z^2_{max}-z^2}}=\nonumber \\
%=\frac{\sqrt{\lambda}}{\pi}
%z_{max}=
&=&\frac{\sqrt{\lambda}}
{\pi}\sqrt{1-(\frac{E^2}{\lambda}-C^2-v_iv^i)v^2} \ .
\nonumber \\
\end{eqnarray}
% where the conserved dilaton charge (\ref{Dcharge}) is equal to
% $D=\sqrt{\lambda}C$ for the ansatz (\ref{ans}).
 As the next step
we calculate the momentum defined (\ref{ptriangle}) for
$z_{min}=0$
\begin{eqnarray}\label{ptriangleinf}
p=2v\omega\int_0^{z_{max}}dz
\frac{z}{(1-z^2)\sqrt{z^2_{max}-
z^2}}=
%\frac{2v\sqrt{\frac{E^2}{\lambda}-C^2}}
%{\sqrt{1-z^2_{max}}}\tan^{-1}\frac{z_{max}}{
%\sqrt{1-z^2_{max}}}=\nonumber \\
%\frac{2v\sqrt{\frac{E^2}{\lambda}-C^2}%}
%{\sqrt{1-z^2_{max}}}\cos^{-1}\sqrt{1-z^2_{max}}=
2\cos^{-1}v\sqrt{\frac{E^2}{\lambda}-C^2-v_iv^i} \ .
\nonumber \\
\end{eqnarray}
%using
%\begin{equation}
%\tan^{-1}(\frac{X}{Y})=
%\cos^{-1}\frac{1}{\sqrt{1+(\frac{X}{Y})^2}}
%\end{equation}
Then finally using (\ref{ptriangleinf}) we can rewrite the
dispersion relation (\ref{di1}) into a more natural form
%Using the result above we can find
%the dependence of $v$ on $p$ as
%\begin{equation}
%v=\frac{\cos\frac{p}{2}}{\sqrt{\frac{E^2}{\lambda}
%-C^2}}
%\end{equation}
\begin{equation}\label{drf}
\sqrt{E^2-D^2-(P_{y_i})^2}-J = \frac{\sqrt{\lambda}}{\pi}
\left|\sin\frac{p}{2}\right| \ .
\end{equation}
A comment is in order for the magnon dispersion relation
(\ref{drf}). We see that there is  a natural symmetry between $D$
and $P_{y_i}$. It is clear that we could generalize the dispersion
relation for giant magnon in $AdS_5\times S^5$ in the same way
when we include the motion of the string along the $AdS_5$
boundary. In other words the presence of the charge $D$ does not
imply any new physical interpretation of the dispersion relation
given above. It simply reflects the fact that the motion of the
string along the radial direction in the near horizon limit of
$NS5$-brane background is free.
%%%%%%%%%%%%%%%%%%%%%%%%%%%%%%%%%%%%%%%%%%
\section{Uniform light-cone gauge}
In this section, we would like to investigate
the giant magnon-like dispersion relation (\ref{drf}) for the open
string orbiting in the background of $N$ NS 5-brane
in more details. For example,
does the giant magnon solution pose uniform distribution of the
angular momentum along the string world-sheet as in case of giant
magnon in $AdS_5\times S^5$ \cite{Arutyunov:2006gs} ? To answer
this question we have to develop the Hamiltonian formalism for
string in the near horizon region of $N$ NS5-branes with general
world-sheet metric. However this requirement makes the analysis
very obscure due to the presence of the non-trivial coupling of
the string to the dilaton. On the other hand when we restrict to
the study of the classical dynamics of string we have to take
$\sqrt{\lambda}\rightarrow \infty$. Then in the first
approximation we can neglect the coupling of the string to the
dilaton when we will try to find the solution in uniform gauge.
Let us proceed along this line.

We again start with the action
\begin{equation}\label{Sun}
S=-\frac{\sqrt{\lambda}}{4\pi}
\int_{-\pi}^\pi d\sigma d\tau
[\sqrt{-\gamma}\gamma^{\alpha\beta}
g_{MN}\partial_\alpha x^M\partial_\beta x^N
-\epsilon^{\alpha\beta}
\partial_\alpha x^M\partial_\beta x^N b_{MN}] \ .
% +\frac{1}{4\pi}
% \int d\sigma d\tau \sqrt{-\gamma}
% R \Phi \ .
\end{equation}
To implement the uniform light-cone
we will follow
\cite{Arutyunov:2004yx}.
Using (\ref{Sun}) we determine
the momenta $p_M$ conjugate to
$x^M$
\begin{equation}
p_M=-\frac{\sqrt{\lambda}}{2\pi}
(\sqrt{-\gamma}\gamma^{\tau\alpha}
g_{MN}\partial_\beta x^N-
b_{MN}\partial_\sigma x^N)
\end{equation}
Let us also use following
parametrization of the metric variables
$\gamma_{\mu\nu}$
\begin{equation}\label{metlambda}
\lambda^\pm=\frac{\sqrt{-\gamma}
\pm \gamma_{\tau\sigma}}{\gamma_{\sigma\sigma}} \ ,
\quad \xi=\ln \gamma_{\sigma\sigma} \ ,
\end{equation}
where $\lambda^\pm$ are manifestly invariant under
the Weyl transformations of metric
\begin{equation}
\gamma'_{\alpha\beta}(\sigma,\tau)=
e^{\phi(\tau,\sigma)}\gamma_{\alpha\beta}
(\tau,\sigma)
\end{equation}
while $\xi$ transforms as $\xi'(\sigma,\tau)=
\xi(\sigma,\tau)+\phi(\sigma,\tau)$. Then since the action does
not contain the time-derivative of the metric we find that the
momenta conjugate to $\lambda^\pm,\xi$ are zero:
\begin{equation}
\pi_\pm=\frac{\delta S}{\delta
\partial_\tau \lambda^\pm}=0 \ ,
\quad \pi_\xi=
\frac{\delta S}{\delta \partial_\tau
\xi}=0 \
\end{equation}
and form the primary constraints of the theory. As the next step
we determine the Hamiltonian density
% and hence
% \begin{equation}
% \partial_\tau x^M=
% -\frac{1}{\sqrt{-\gamma}\gamma^{\tau\tau}}
% [\frac{2\pi}{\sqrt{\lambda}}p_N
% +\partial_\sigma x^K b_{KN}+
% \sqrt{-\gamma}\gamma^{\tau\sigma}\partial_\sigma
% x^K g_{KN}]g^{NM}
% \end{equation}
\begin{eqnarray}
\mH_0 &=& \pi_+\partial_\tau
\lambda^++\pi_-\partial_\tau \lambda^-
+\pi_\xi \partial_\tau \xi
+\partial_\tau x^Mp_M-\mL=\nonumber \\
&=&
% \partial_\tau
% \gamma_{\alpha\beta}
% \pi^{\alpha\beta}
-\frac{1}{\sqrt{-\gamma}
\gamma^{\tau\tau}}
[\frac{\pi}{\sqrt{\lambda}}
p_M g^{MN}p_N+\frac{\sqrt{\lambda}}{4\pi}
g_{MN}\partial_\sigma x^M\partial_\sigma x^N
+p_M g^{MN}b_{NK}\partial_\sigma x^K+
\nonumber \\
&+&\frac{\sqrt{\lambda}}{4\pi}
\partial_\sigma x^M b_{MN}g^{NK}
b_{KL}\partial_\sigma x^L]
-\frac{\gamma^{\tau\sigma}}{\gamma^{\tau\tau}}
p_M\partial_\sigma x^M
% -\frac{1}{4\pi}
% \sqrt{-\gamma}R \Phi
% =
% \nonumber \\
% =-\frac{1}{\sqrt{-\gamma}
% \gamma^{\tau\tau}}[
% \frac{\pi}{\sqrt{\lambda}}
% \Pi_M g^{MN}\Pi_N+
% \frac{\sqrt{\lambda}}{2\pi}
% \partial_\sigma x^M g_{MN}\partial_\sigma x^N]-
% \frac{\gamma^{\tau\sigma}}{\gamma^{\tau\tau}}
% \Pi_M\partial_\sigma x^M+\nonumber \\
% \partial_\tau
% \gamma_{\alpha\beta}
% \pi^{\alpha\beta}-\frac{1}{4\pi}
% \sqrt{-\gamma}R \Phi \ , \quad
% \Pi_M=p_M+\frac{\sqrt{\lambda}}{2\pi}
% b_{MN}\partial_\sigma x^N \ ,
 \ ,  \nonumber \\
\end{eqnarray}
or alternatively using the variables
(\ref{metlambda})
\begin{eqnarray}
\mH_0&=&\frac{\lambda^++\lambda^-}{2}
T_0+\frac{\lambda^+-\lambda^-}{2}T_1=
\lambda^+T_++\lambda^-T_- \ , \nonumber \\
T_+&=&\frac{1}{2}(T_0+T_1)\ , \quad
T_-=\frac{1}{2}(T_0-T_1) \ ,  \nonumber \\
\end{eqnarray}
where
\begin{eqnarray}
T_0&=&
[\frac{\pi}{\sqrt{\lambda}}
p_M g^{MN}p_N+\frac{\sqrt{\lambda}}{4\pi}
g_{MN}\partial_\sigma x^M\partial_\sigma x^N
+p_M g^{MN}b_{NK}\partial_\sigma x^K+
\nonumber \\
&+&\frac{\sqrt{\lambda}}{4\pi}
\partial_\sigma x^M b_{MN}g^{NK}
b_{KL}\partial_\sigma x^L] \ , \nonumber \\
T_1&=&\partial_\sigma x^M p_M \ .
\nonumber \\
\end{eqnarray}
The stability of the primary constraints
imply the secondary ones
\cite{Henneaux:1992ig,Govaerts:1991gd,Govaerts:2002fq}
\begin{equation}
T_0=T_1=0 \ .
\end{equation}
It can be shown that stability of these constraints does not imply
any additional ones.

Note also that the invariance of the
action (\ref{Sun}) under
the shift $\phi_2'=\phi_2+\epsilon$
implies an existence of the conserved
quantity
\begin{equation}
P_{\phi_2}=\frac{\sqrt{\lambda}}
{2\pi}\int_{-\pi}^{\pi}
d\sigma [\sqrt{-\gamma}\gamma^{\tau
\alpha}g_{\phi_2 N}\partial_\alpha x^N
-\epsilon^{\tau\alpha}b_{\phi_2 N}\partial_\alpha
x^N]=-\int_{-\pi}^\pi d\sigma
p_{\phi_2} \ .
\end{equation}
Our goal is to develop the Hamiltonian formulation of completely
fixed gauge theory. Note that the momenta $\pi_\pm$ and $\pi_\xi$
decouple from the theory and these constraints can be ignored.
Then we can interpret $\lambda^\pm$ as the Lagrange multipliers
for the constraints $T_\pm$. We fix the gauge symmetry generated
by $T_0,T_1$ if we impose following uniform light-cone gauge
\cite{Arutyunov:2004yx}
\begin{equation}\label{gaugegix}
p_{\phi_2}=\frac{1}{2\pi}J \ , \quad
t= \tau \ .
 \end{equation}
The fixing the gauge implies
 that the original Hamiltonian
$H_0=\int_{-\pi}^\pi
d\sigma \mH_0$ strongly vanishes.
Secondly, the stability of the
gauge fixing functions (\ref{gaugegix})
determine the values of $\lambda_\pm$.
However the explicit values of these
constraints are not important for
our purposes.

The fact that $T_0,T_1$
strongly vanish allows us to find the dynamics
on the reduced phase space. Firstly,
from $T_1$ we express $\partial_\sigma \phi_2$
as
\begin{equation}
\partial_\sigma \phi_2=-\frac{1}{p_{\phi_2}}
\partial_\sigma x^m p_m \ ,
\end{equation}
where $m$ now label all coordinates different from $t,\phi_2$.
Note that $x^m,p_m$ parameterize the reduced phase space.

Secondly, the gauge fixing (\ref{gaugegix}) implies that the
Hamiltonian that governs the dynamics on the reduced phase space
should be identified with $-p_t$. This follows from the fact that
for the gauge fixing (\ref{gaugegix})  the action takes the form
\begin{equation}
S=\int d\tau d\sigma
[\dot{x}^mp_m+p_t]
\end{equation}
that  suggests that the Hamiltonian
density on reduced phase space
$\mH$ should be identified with
$-p_t$. Then the
time evolution of  $p_m,x^m$ is governed
by following equation
\begin{equation}
\partial_\tau p_m=\pb{p_m,H}_D \ , \quad
\partial_\tau x^m=\pb{x^m,H}_D \ ,
 \end{equation}
 where $H=\int_{-\pi}^\pi
 d\sigma \mH$ and
 where subscript $D$ means Dirac bracket that
however for the gauge fixing (\ref{gaugegix})
coincides with the original Poisson brackets
\begin{equation}
\pb{x^m(\sigma),p_n(\sigma')}=
\delta_n^m\delta(\sigma-\sigma') \ .
\end{equation}
As the next step we express $p_t$ as
 functions of the canonical
variables $p_m,x^m$ with the help of $T_1$ and $T_0$
\begin{eqnarray}
p_t^2
% =\frac{\sqrt{\lambda}}{\pi}
% (\frac{\pi}{\sqrt{\lambda}}
% p_m g^{mn}p_n+\frac{\pi}{\sqrt{\lambda}}
% p_{\phi_2}^2 g^{\phi_2\phi_2}+
% \frac{\sqrt{\lambda}}
% {4\pi}\partial_\sigma x^m g_{mn}
% \partial_\sigma x^n+\frac{\sqrt{\lambda}}
% {4\pi}\frac{1}{p_\phi^2}g_{\phi_2\phi_2}
% (p_m\partial_\sigma x^m)^2+
% \nonumber \\
% +p_\theta g^{\theta\theta}
% b_{\theta\phi_2}\partial_\sigma \phi^2+
% p_{\phi_2}g^{\phi_2\phi_2}b_{\phi_2
% \theta}\partial_\sigma \theta
% +\frac{\sqrt{\lambda}}
% {4\pi}
% \partial_\sigma \theta b_{\theta\phi_2}
% g^{\phi_2\phi_2}b_{\phi_2\theta}\partial_\sigma
% \theta
% +\frac{\sqrt{\lambda}}
% {4\pi}
% \partial_\sigma  \phi_2 b_{\phi_2\theta}
% g^{\theta\theta}b_{\theta\phi_2}
% \partial_\sigma \phi_2)=
% \nonumber \\
&=&\frac{\sqrt{\lambda}}{\pi}
(\frac{\pi}{\sqrt{\lambda}}
p_m g^{mn}p_n+\frac{\pi}{\sqrt{\lambda}}
p_{\phi_2}^2 g^{\phi_2\phi_2}+
\frac{\sqrt{\lambda}}
{4\pi}\partial_\sigma x^m g_{mn}
\partial_\sigma x^n+\frac{\sqrt{\lambda}}
{4\pi}\frac{g_{\phi_2\phi_2}}{p_{\phi_2}^2}
(p_m\partial_\sigma x^m)^2-
\nonumber \\
&-& p_\theta g^{\theta\theta}
b_{\theta\phi_2}\frac{1}{p_{\phi_2}}
\partial_\sigma x^mp_m+
p_{\phi_2}g^{\phi_2\phi_2}b_{\phi_2
\theta}\partial_\sigma \theta
+\nonumber \\
&+&\frac{\sqrt{\lambda}}
{4\pi}
\partial_\sigma \theta b_{\theta\phi_2}
g^{\phi_2\phi_2}b_{\phi_2\theta}\partial_\sigma
\theta
+\frac{\sqrt{\lambda}}
{4\pi p_{\phi_2}^2}
(p_m \partial_\sigma x^m) b_{\phi_2\theta}
g^{\theta\theta}b_{\theta\phi_2}
(p_m\partial_\sigma x^m))\equiv \bK \ .
\nonumber \\
\end{eqnarray}
Then the Hamiltonian density $\mH$ takes
the form
\begin{equation}
\mH=\sqrt{\bK} \ ,
\end{equation}
where we have picked up the negative
root in the equation $p_t^2=\bK$
\cite{Arutyunov:2004yx}.

To simplify the analysis further note that
 five coordinates $y^i,i=0,\dots,5$
and the  corresponding momenta $p_{y_i}$
that  determine the dynamics of the
 string along the world-volume
of NS5-branes  are free.
 Then we can solve their equations of
motion by requirement that $y^i$
are constant and $p_{y_i}=0$.
Then the remaining degrees of freedom are
\begin{equation}
(r,p_r),\quad (\theta,p_\theta),
\quad (\phi_1,p_{\phi_1})
\end{equation}
and we presume that they spatial and
time dependence  has following
form
\begin{eqnarray}\label{ansh}
r&=&r(\tau), \quad p_r=p_r(\tau), \nonumber \\
\phi_1&=&\phi_1^{const}, \quad p_{\phi_1}=0 \ ,
\nonumber \\
\theta&=&\theta(\sigma-v\tau), \quad
p_\theta=p_\theta(\sigma-v\tau) \ .  \nonumber \\
\end{eqnarray}
% Furhter, since the Hamiltonian is conserved
% we can demand that
% \begin{equation}
% H=E=\int_0^{2\pi} d\sigma \mH
% \end{equation}
% and the conservation of $H$ implies
% \begin{equation}
% \partial_\tau \mH=
% \partial_\sigma \mH_1
% \end{equation}
% for some function $\mH_1$.
First of all let us determine the values
of $\phi_1$. The consistency of
 the ansatz (\ref{ansh})
implies that
% The equation of motion for
% $\phi_1,p_{\phi_1}$ take the form
% \begin{eqnarray}
% \partial_\tau
% \phi_1&=&0 \ , \nonumber \\
% \partial_\tau p_{\phi_1}&=&
% -[-\frac{p^2_{\phi_2}\cosh\phi_1}{
% \sin^2\theta \sin^3\phi_1}
% +\frac{\sqrt{\lambda}}{4\pi^2}
% \sin^2\theta \sin\phi_1\cos\phi_1
% (p_m\partial_\sigma x^m)+\nonumber \\
% &+&2p_{\theta}g^{\theta\theta}
% \sin^2\theta \cos\phi_1\sin\phi_1
% (\partial_\sigma x^m p_m)-\nonumber \\
% &-& 2p_{\phi_2}\frac{\sin\phi_1\cos\phi_1}
% {\sin^2\theta \sin^3\phi_{1}}
% \sin^2\theta\cos^2\phi_1\partial_\sigma\theta
% -2p_{\phi_2}\frac{1}{\sin^2\theta
% \sin^2\phi_1}\sin\phi_1\cos\phi_1
% \partial_\sigma \theta]
% \frac{1}{\sqrt{\bK}} \ , \nonumber \\
% \end{eqnarray}
% where we have inserted the ansatz $p_{\phi_1}=0 \ ,
% \phi_1=\mathrm{const}$. The requrament that
% $\partial_\tau p_{\phi_1}=0$ implies
$\phi_1=\frac{\pi}{2}$. However for this
value the NS two form field
(\ref{NS5bacd})
vanishes. In what follows
we use the notation $g_{\phi_2\phi_2}
\equiv g_{\phi_2\phi_2}(\phi_1=\pi/2)=
\sin^2\theta$.
Further, the equations of motion for $p_r,r$ take
the form
\begin{equation}\label{eqr}
\partial_\tau r=
\frac{p_rr^2}{\sqrt{\bK}} \ , \quad
\partial_\tau p_r=-
\frac{p_r^2 r}{\sqrt{\bK}} \
\end{equation}
that imply
\begin{equation}
\partial_\tau (p_r^2r^2)=0 \ .
% \frac{2}{
% \sqrt{\bK}}
% \left(p_r^3r^2-p_r^3r^2
% \right)=0
\end{equation}
In other words the expression $p_r^2 g^{rr}$ that
is presented in $\bK$ is constant.
% In following
% we  denote this constant as
%   $C'$.
Then the function $\bK$
takes the form
\begin{eqnarray}
\bK=\frac{\sqrt{\lambda}}{\pi}
(\frac{\pi}{\sqrt{\lambda}}p_r^2 g^{rr}
+\frac{\pi}{\sqrt{\lambda}}
p_{\phi_2}^2 g^{\phi_2\phi_2}+
\frac{\pi}{\sqrt{\lambda}}p_\theta^2g^{\theta\theta}+
\frac{\sqrt{\lambda}}
{4\pi}\partial_\sigma \theta'^2 g_{\theta\theta}
+\frac{\sqrt{\lambda}}
{4\pi}\frac{g_{\phi_2\phi_2}}{p_{\phi_2}^2}
(p_\theta\partial_\sigma \theta)^2) \ ,
\nonumber \\
% +(p_\theta^2 g^{\theta\theta}+
% p_{\phi_2}^2g^{\phi_2\phi_2})
% b_{\phi_2\theta}\frac{1}{p_{\phi_2}}
% \partial_\sigma \theta+
% \frac{\sqrt{\lambda}}
% {4\pi}(p_{\phi_2}^2g^{\phi_2\phi_2}+
% p_{\theta}^2g^{\theta\theta})
% \frac{1}{p_{\phi_2}^2}  b_{\theta\phi_2}
% b_{\phi_2\theta}\partial_\sigma
% \theta^2 \nonumber \\
\end{eqnarray}
where we have used the fact that $b_{\phi_2\theta}(\phi_1
=\frac{\pi}{2})=0$.
 Note also that
the equation of motion for $\theta$ takes the form
\begin{eqnarray}\label{eqtheta}
\partial_\tau \theta=
\frac{1}{\sqrt{\bK}}
(p_\theta g^{\theta\theta}
+\frac{\lambda}{4\pi^2}
\frac{g_{\phi_2\phi_2}}{p_{\phi_2}^2}
(\partial_\sigma \theta p_\theta
\partial_\sigma \theta))
\nonumber \\
% -g^{\theta\theta}b_{\theta\phi_2}\frac{1}{p_{\phi_2}}
% \partial_\sigma \theta p_\theta
%  +
% \frac{\sqrt{\lambda}}{4\pi p_{\phi_2}^2}
% \partial_\sigma \theta b_{\phi_2
% \theta}g^{\theta\theta}b_{\theta\phi_2}
% (p_\theta\partial_\sigma \theta))
% \nonumber \\
\end{eqnarray}
Since the problem reduces to the
study of the dynamics of $\theta$ and $r$ only
it is more natural to proceed to the Lagrangian
formalism. Using (\ref{eqr}) and (\ref{eqtheta})
and after some algebra we find the Lagrangian
density defined as
\begin{eqnarray}
\mL=\partial_\tau r p_r+
\partial_\tau\theta p_\theta-\sqrt{\bK}
% =\nonumber \\
% -\frac{\sqrt{\lambda}}{\pi
% \sqrt{\bK}}
% [\frac{\pi}{\sqrt{\lambda}}
% p_{\phi_2}^2 g^{\phi_2\phi_2}+
% \frac{\sqrt{\lambda}}{4\pi}
% \partial_\sigma \theta g_{\theta\theta}
% \partial_\sigma \theta]
% +p_{\phi_2}
% g^{\phi_2\phi_2}b_{\phi_2\theta}
% \partial_\sigma \theta+
% \frac{\sqrt{\lambda}}{4\pi}
% \partial_\sigma \theta b_{\theta\phi_2}
% g^{\phi_2\phi_2}b_{\phi_2\theta}
% \partial_\sigma \theta ]
\nonumber \\
\end{eqnarray}
% Let us write schematically $\bK=\frac{\sqrt{\lambda}}
% {\pi}[\frac{\pi}{\sqrt{\lambda}}
% (p_r^2g^{rr}+Ep_\theta^2)+A]$.
% Then we have
% \begin{eqnarray}
% (\frac{\dot{\theta}^2}{E}g_{\theta\theta}+
% \dot{r}^2 g_{rr})
%(\frac{\pi}{\sqrt{\lambda}})^2
% \bK=p_r^2 g^{rr}+p_\theta^2 E \Rightarrow
% \nonumber \\
% (p_r^2 g^{rr}+p_\theta^2 E)=
% \frac{(\frac{\dot{\theta}^2}{E}g_{\theta\theta}+
% \dot{r}^2 g_{rr})\frac{\sqrt{\lambda}}
% {\pi} A}
% {1-(\frac{\dot{\theta}^2}{E}g_{\theta\theta}+
% \dot{r}^2 g_{rr})
%(\frac{\pi}{\sqrt{\lambda}})^2
% }\Rightarrow
% \nonumber \\
% \bK=\frac{\sqrt{\lambda}}{\pi}
% \frac{A}{1-(\frac{\dot{\theta}^2}{E}g_{\theta\theta}+
% \dot{r}^2 g_{rr})
%(\frac{\pi}{\sqrt{\lambda}})^2
% }\nonumber \\
% \end{eqnarray}
takes the form
\begin{eqnarray}\label{mLred}
\mL=
% -\sqrt{\frac{\sqrt{\lambda}}{\pi}
% (\frac{\pi}{\sqrt{\lambda}}p_{\phi_2}^2
% g^{\phi_2\phi_2}+\frac{\sqrt{\lambda}}{4\pi}
% \partial_\sigma \theta g^{\theta\theta}
% \partial_\sigma \theta}
% +p_{\phi_2}
% g^{\phi_2\phi_2}b_{\phi_2\theta}
% \partial_\sigma \theta+
% \frac{\sqrt{\lambda}}{4\pi}
% \partial_\sigma \theta b_{\theta\phi_2}
% g^{\phi_2\phi_2}b_{\phi_2\theta}
% \partial_\sigma \theta }
% \times
% \sqrt{1-(\frac{\dot{\theta}^2}{E(\theta)}g_{\theta\theta}+
% \dot{r}^2 g_{rr})}= \nonumber \\
-p_{\phi_2}
\sqrt{[g^{\phi_2\phi_2}
+\frac{\lambda}{4\pi^2}
\frac{1}{p_{\phi_2}^2}
(\partial_\sigma \theta)^2
% \frac{\sqrt{\lambda}}{\pi}
% g^{\phi_2\phi_2}
% b_{\phi_2\theta}\frac{1}{p_{\phi_2}}
% \partial_\sigma \theta
%  +
% \frac{\lambda}{4\pi^2 p_{\phi_2}^2}
% \partial_\sigma \theta g^{\phi_2\phi_2}
% b_{\phi_2
% \theta}b_{\theta\phi_2}
% \partial_\sigma \theta
][1-(\partial_\tau r)^2g_{rr}]
-g^{\phi_2\phi_2}g_{\theta\theta}^2
(\partial_\tau \theta)^2} \ .
\nonumber \\
\end{eqnarray}
% where
% \begin{eqnarray}
% E= g^{\theta\theta}
% +\frac{\sqrt{\lambda}}{\pi}
% \frac{\sqrt{\lambda}}{4\pi}
% \frac{g_{\phi_2\phi_2}}{p_{\phi_2}^2}
% \partial_\sigma \theta
% \partial_\sigma \theta-
% \nonumber \\
% -\frac{\sqrt{\lambda}}{\pi}
% g^{\theta\theta}b_{\theta\phi_2}\frac{1}{p_{\phi_2}}
% \partial_\sigma \theta
%  +
% \frac{\sqrt{\lambda}}{\pi}
% \frac{\sqrt{\lambda}}{4\pi p_{\phi_2}^2}
% \partial_\sigma \theta b_{\phi_2
% \theta}g^{\theta\theta}b_{\theta\phi_2}
% \partial_\sigma \theta \Rightarrow
% \nonumber \\
% \frac{\pi}{\sqrt{\lambda}}
% \frac{p_{\phi_2}^2g^{\phi_2\phi_2}}
% {g^{\theta\theta}}E=
% \frac{\pi}{\sqrt{\lambda}}
% g^{\phi_2\phi_2}p_{\phi_2}^2+
% \frac{\sqrt{\lambda}}{4\pi}g_{\theta\theta}
% (\partial_\sigma \theta)^2+
% \nonumber \\
% +p_{\phi_2}b_{\phi_2\theta}g^{\phi_2\phi_2}
% \partial_\sigma \theta+
% \frac{\sqrt{\lambda}}{4\pi }
% \partial_\sigma \theta b_{\theta \phi_2}
% g^{\theta\theta}b_{\theta\phi_2}
% \partial_\sigma \theta=A
% \nonumber \\
% \end{eqnarray}
As the next step we insert the ansatz
$\theta=\theta(\sigma-v\tau)$ into (\ref{mLred}).
Since
$\dot{R}=C'R$ we obtain that (\ref{mLred})
 takes the form
\begin{eqnarray}\label{mLu}
\mL_{red}
% =-p_{\phi_2}\times \nonumber \\
% \times
% \sqrt{(g^{\phi_2\phi_2}
% +\frac{\lambda}{4\pi^2}
% \frac{1}{p_{\phi_2}^2}
% \theta'^2
% +\frac{\sqrt{\lambda}}{\pi}g^{\phi_2\phi_2}
% b_{\phi_2\theta}\frac{1}{p_{\phi_2}}
% \theta'
%  +
% \frac{\lambda}{4\pi^2 p_{\phi_2}^2}
%  g^{\phi_2\phi_2} b_{\phi_2
% \theta}b_{\theta\phi_2}
% \theta'^2
% )(1-C'^2)
%  -g^{\phi_2\phi_2}v^2\theta'^2
% }=
% \nonumber \\
% =-\sqrt{(p_{\phi_2}^2g^{\phi_2\phi_2}
% +\frac{\lambda}{4\pi^2}
% \theta'^2
%+% \frac{p_{\phi_2}\sqrt{\lambda}}{\pi}
% g^{\phi_2\phi_2}b_{\phi_2\theta}
% \theta'
%  +
% \frac{\lambda}{4\pi^2}
% g^{\phi_2\phi_2}  b_{\phi_2
% \theta}b_{\theta\phi_2}
% \theta'^2
% )(1-C'^2)
% -p_{\phi_2}^2
% g^{\phi_2\phi_2}v^2\theta'^2}=
% \nonumber \\
=-
\sqrt{g^{\phi_2\phi_2}p_{\phi_2}^2
(1-C'^2)-
(p^2_{\phi_2}
v^2g^{\phi_2\phi_2}-(1-C'^2)\frac{\lambda}{4\pi^2}
% +\frac{\lambda}{\pi^2}
% g_{\phi_2\phi_2})
)\theta'^2
% -(1-C'^2)\frac{2\sqrt{\lambda}}{\pi}
% \theta'}
} \ .
\nonumber \\
\end{eqnarray}
% using the fact that
% \begin{equation}
% b_{\theta\phi_2}=2g_{\phi_2
% \phi_2} \ .
% \end{equation}
In other words the problem reduces
 to the dynamics  of one degree
of freedom $\theta$ that depends on the parameter
$u\equiv\sigma-v\tau$. Then
 it is convenient
to find the
corresponding
 Hamiltonian since using the fact that
(\ref{mLu})
does not explicitly depend
 on $u$ the corresponding
Hamiltonian,
 that determines the
 evolution with respect to the
parameter $u$ is  conserved.
 Explicitly, the conjugate momentum
to $\theta$ takes the form
\begin{equation}
\bp=
\frac{\delta \mL}{\delta \theta'}=
\frac{(p^2_{\phi_2}
v^2g^{\phi_2\phi_2}+(C'^2-1)\frac{\lambda}{4\pi^2}
)\theta'}{\sqrt{(\dots)}}
\end{equation}
and hence the reduced Hamiltonian is equal to
\begin{eqnarray}\label{Hred}
H^{red}=\theta'\bp-\mL^{red}=
\frac{g^{\phi_2\phi_2}
p_{\phi_2}^2(1-C'^2)}
{\sqrt{(\dots)}} \ .
\nonumber \\
\end{eqnarray}
Since the reduced Hamiltonian obeys
$\frac{dH^{red}}{du}=0$ we set it to
some constant
\begin{equation}
H^{red}=R \ .
\end{equation}
Then from (\ref{Hred}) we can find
the differential equation for $\theta'$
\begin{eqnarray} \label{theta'}
\theta'^2=
% \frac{g^{\phi_2\phi_2}p^2_{\phi_2}
% (1-C'^2)(R^2-g^{\phi_2\phi_2}
% p^2_{\phi_2}(1-C'^2))}
% {R^2(p^2_{\phi_2}vg^{\phi_2
% \phi_2}-(1-C'^2)\frac{\lambda}{4\pi^2})}=
% \nonumber \\
\frac{4\pi^2p^2_{\phi_2}}
{\lambda \sin^2\theta}
\frac{\sin^2\theta-\frac{p^2_{\phi_2}(1-C'^2)}
{R^2}}{
\frac{4\pi^2 p^2_{\phi_2}v^2}
{\lambda(1-C'^2)}-\sin^2\theta} \ .
\nonumber \\
\end{eqnarray}
The equation (\ref{theta'})
implies following bound on the allowed
values of $\theta$
\begin{eqnarray}
\frac{p^2_{\phi}(1-C'^2)}{R} < \sin^2\theta
<  \frac{4\pi^2}{\lambda}
\frac{p_{\phi_2}^2v^2}{1-C'^2} \ .
\nonumber \\
\end{eqnarray}
It is again convenient to introduce the
 variable
$z^2=1-\sin^2\theta$ that now obeys
\begin{eqnarray}\label{zeqD}
z'^2&=&
% (1-z^2)\theta'^2=
\frac{4\pi^2p^2_{\phi_2}}
{\lambda }
\frac{z^2-z^2_{min}}{z_{max}^2-z^2}
\ , \quad  z_{min}\leq z \leq z_{max} \ ,
\nonumber \\
z_{min}^2&=&1-\frac{p^2_{\phi_2}(1-C'^2)}{R^2} \ ,
\quad
z_{max}^2=1-\frac{4\pi^2}{\lambda}\frac{p^2_{\phi_2}
v^2}{(1-C'^2)} \ .
\nonumber \\
\end{eqnarray}
 We see from (\ref{zeqD}) that the
case of infinite giant magnon corresponding
to $z_{min}=0$  occurs
for  $R^2=p^2_{\phi_2}(1-C'^2)$.

Let us now  calculate $p_t$ for this case.
Since $p_t=-\sqrt{\bK}$ we obtain
\begin{eqnarray}
p_t=-
% \sqrt{\frac{p^2_{\phi_2}g^{\phi_2\phi_2}}
% {g^{\theta\theta}}
% \frac{(g^{\theta\theta}+\frac{\lambda}{4\pi^2}
% \frac{g_{\phi_2\phi_2}}{p^2_{\phi_2}}(\partial_\sigma
% \theta)^2)^2}{
% (g^{\theta\theta}+\frac{\lambda}{4\pi^2}
% \frac{g_{\phi_2\phi_2}}{p^2_{\phi_2}}(\partial_\sigma
% \theta)^2)(1-(\partial_\tau r)^2g_{rr})-(\partial_\tau
% \theta)^2g_{\theta\theta}}}=
% \nonumber \\
%  =\frac{p_{\phi_2}}{\sin\theta}
%  \frac{1+\frac{z^2-z^2_{min}}
%  {z^2_{max}-z^2}}
%  {\sqrt{(1+
%  \frac{z^2-z^2_{min}}{z^2_{max}-z^2})
%  (1-C'^2)
%  -\frac{4\pi^2}{\lambda}
%  \frac{p^2_{\phi_2}v^2}{(1-z^2)}
%  \frac{z^2-z^2_{min}}{z^2_{max}-z^2}}}=
% \nonumber \\
% =p_{\phi_2}\frac{z^2_{max}-z^2_{min}}
% {\sqrt{z^2_{max}-z^2}\sqrt{1-C'^2}}
% \frac{1}{
% \sqrt{(z^2_{max}-z^2_{min})
% (1-z^2)+(z^2_{max}-1)
% (z^2-z^2_{min})}}=
% \nonumber \\
%\sqrt{z^2_{max}((1-C'^2)-\frac{z^2_{min}}
%{1-C'^2})+z^2
%(z^2_{max}-1+z^2_{min}(1-C'^2))}}
%\nonumber \\
% =p_{\phi_2}\frac{z^2_{max}-z^2_{min}}
% {\sqrt{z^2_{max}-z^2}\sqrt{1-C'^2}}
% \frac{1}{
% \sqrt{(z^2_{max}-z^2)(1-z^2_{min})}}=
% \nonumber \\
\frac{R}{(1-C'^2)}
\frac{z^2_{max}-z^2_{min}}
{\sqrt{z^2_{max}-z^2}}
\frac{1}{\sqrt{(z^2_{max}-z^2)}}
\nonumber \\
\end{eqnarray}
and hence
\begin{equation}
\frac{p_t}{|z'|}=
-\frac{\sqrt{\lambda}}{2\pi }
\frac{R}{(1-C'^2)p_{\phi_2}}
\frac{z^2_{max}-z^2_{min}}
{\sqrt{(z^2_{max}-z^2)(z^2-z^2_{min})}
} \ .
\end{equation}
Now it turns out that the dispersion relation  has following form
% The important question is what disperrsion
% relation we should expect.
% To find let us again restrict to the case
% of infinite giant with $z_{min}^2=1$. Then
% in order to find the finite contribution
% from the energy given above we have to
% extract the constant part there.
% Let us then presume an existence of following
% quantity
% \begin{equation}
% K=\frac{K}{2\pi}
% \int_{-\pi}^\pi d\sigma=
% 2\frac{K}{2\pi}\int_{z_{min}}^{z_{max}}
% \frac{dz}{|z'|}=
% \frac{\sqrt{\lambda}}
% {2\pi^2}\frac{K}{p_{\phi_2}}
% \int_{z_{min}}^{z_{max}}
% dz\frac{\sqrt{z^2_{max}-z^2}}
% {\sqrt{z^2-z^2_{min}}}
% \end{equation}
\begin{eqnarray}\label{disrelu}
E-\frac{J}{\sqrt{1-C'^2}}
% =-
% \frac{\sqrt{\lambda}}
% {2\pi(1-C'^2)p_{\phi_2}}
% \int_{z_{min}}^{z_{max}}
% dz \frac{ z^2_{max}(2R+\frac{K}{\pi}(1-C'^2))
% -\frac{K}{\pi}z^2(1-C'^2)}
% {\sqrt{(z^2_{max}-z^2)}z}
% \Rightarrow
% \nonumber \\
% K=-2\pi \frac{p_{\phi_2}}{\sqrt{1-C'^2}}=
% -\frac{J}{\sqrt{1-C'^2}}
%  \ , \Rightarrow
% \nonumber \\
% E-K
=\frac{\sqrt{\lambda}}{\pi}
\frac{1}{\sqrt{1-C'^2}}
\sqrt{1-\frac{4\pi^2 p^2_{\phi_2}v^2}
{\lambda(1-C'^2)}}=
\frac{1}{\sqrt{1-C'^2}}
\frac{\sqrt{\lambda}}{\pi}
|\sin \frac{p}{2}|
\ \nonumber \\
\end{eqnarray}
using the fact that
\begin{eqnarray}
p=\triangle \phi_2=
 \int_{-\pi}^\pi d\sigma
 \partial_\sigma \phi_2=
 -\frac{1}{p_{\phi_2}}
 \int_{-\pi}^\pi d\sigma
 p_{\theta}\partial_\sigma
 \theta=
% \nonumber \\
% -\frac{2}{p_{\phi_2}}
% \int_{\theta_{min}}^{\theta_{max}}
% d\theta \frac{p_\theta\partial_\sigma
% \theta}{|\theta'|}=
% \frac{2}{p_{\phi_2}}
% \int_{\theta_{min}}^{\theta_{max}}
% d\theta |p_\theta|=
% \nonumber \\
% =\frac{2}{p_{\phi_2}}
% \int_{\theta_{min}}^{\theta_{max}}d\theta
% \frac{|\partial_\tau \theta \sqrt{\bK}|}
% {1+\sin^2\theta\frac{\lambda}{4\pi^2}
% (\partial_\sigma\theta)^2}=
% =\frac{2v}{p_{\phi_2}}
% \int_{\theta_{min}}^{\theta_{max}} d\theta
% \frac{|\theta'p_t|}
% {1+\sin^2\theta\frac{\lambda}{4\pi^2}
% (\theta')^2}=
% \nonumber \\
\nonumber \\
% =\frac{2v}{p_{\phi_2}}
% \int_{z_{min}}^{z_{max}} \frac{dz}{1-z^2}
% \frac{R}{(1-C^2)}\frac{z^2_{max}-z^2_{min}}
% {\sqrt{(z^2_{max}-z^2)^2}}
% \frac{2\pi p_{\phi_2}}
% {\sqrt{\lambda}}
% \frac{1}
% {1+\frac{z^2-z^2_{min}}{
% z^2_{max}-z^2}}
% \frac{\sqrt{z^2-z^2_{min}}}
% {\sqrt{z^2_{max}-z^2}}=
% \nonumber \\
=\frac{2\pi v}{p_{\phi_2}}
\frac{R}{(1-C^2)}
\int_{z_{min}}^{z_{max}} \frac{dz}{1-z^2}
\frac{2\pi p_{\phi_2}}
{\sqrt{\lambda}}\frac{\sqrt{z^2-z^2_{min}}}
{\sqrt{z^2_{max}-z^2}}
\nonumber \\
% =\frac{4\pi v p_{\phi_2}}{
% \sqrt{\lambda}\sqrt{1-C^2}}
% \int_0^{z_{max}}
% \frac{dz}{1-z^2}\frac{z}{\sqrt{z^2_{max}-z^2}}=
% \nonumber \\
% =\frac{2\pi p_{\phi_2}}
% {\sqrt{\lambda}}\frac{2v}{\sqrt{1-C^2}}
% \frac{1}{\sqrt{1-z^2_{max}}}
% \cos^{-1}\sqrt{1-z^2_{max}}
% \Rightarrow
\end{eqnarray}
that for $z_{min}=0$ implies
\begin{equation}
% p=
% 2
% \cos^{-1}\frac{2\pi}{\sqrt{\lambda}}
% \frac{p_{\phi_2}v}{\sqrt{1-C^2}}
% \nonumber \\
\cos \frac{p}{2}=
\frac{2\pi}{\sqrt{\lambda}}\frac{p_{\phi_2}v}
{\sqrt{1-C^2}} \ .
\end{equation}
Note that we have also presumed
 $-p_\theta \partial_\sigma
\theta'>0$ so that $\partial_\sigma \theta'<0$.
We have to find the relation between
 $C'$ and the conserved dilaton
charge defined in
(\ref{Dchargerho}). To do this
note that the equation of motion for
$r$ is
\begin{equation}\label{eqrpr}
\partial_\tau r=\frac{g^{rr}p_r}{\sqrt{\bK}}
\end{equation}
Note that  we have defined $C'$ through
the relation
$\partial_\tau r=C'r$.  Then (\ref{eqrpr})
implies
\begin{equation}\label{prr}
p_r r=C'\sqrt{\bK}=-C'p_t \ .
\end{equation}
On the other hand the dilaton charge is
equal to
\begin{equation}
D=\int_{-\pi}^{\pi} d\sigma
p_r r \
\end{equation}
so that  using (\ref{prr}) we finally obtain
\begin{equation}
D=C'E \ .
\end{equation}
Then we can rewrite the relation
(\ref{disrelu}) in more suggestive form
%  Using this fact
% and if we denote $\frac{\sqrt{\lambda}}{\pi}
% |\sin\frac{p}{2}|=M$
% we can write the dispression relation given above as
\begin{eqnarray}
\sqrt{E^2-D^2}-J=\frac{\sqrt{\lambda}}{\pi}
|\sin \frac{p}{2}|
\nonumber \\
\end{eqnarray}
that coincides with the result given
in (\ref{drf}) for $P_{y_i}=0$.

\section{Conclusions}
In this paper we have studied the giant magnon solution in the
near horizon limit of NS 5-branes. We derive the magnon like
dispersion relation in Polyakov action of the string in the
conformal gauge. For large $J$ value, we have been able to compute
the one magnon dispersion relation in this background. We further
have examined the magnon solution in the Hamiltonian formalism in
the uniform gauge. The presence of the nontrivial dilaton makes
the formalism obscure, but one can show that there in the limit of
$\sqrt{\lambda} \rightarrow \infty$, the term in the action which
determines the coupling of dilaton becomes very small, and one can
ignore this. By making use of the Hamiltonian formalism we have
been able to reduce it to the dynamics of one degree of freedom.
We have been able to find out the magnon dispersion relation which
matches exactly with the one we find in the Polyakov action.
Moreover, we have argued that this dispersion relation takes the
same form as the dispersion relation in case of giant magnon in
$AdS_5\times S^5$ when we consider the motion along the boundary
of $AdS_5$. This follows from the fact that the motion of the
fundamental string in the near horizon region of $NS5$-branes is
free and it is a reflection of fact that the configuration of the
NS5-brane and fundamental string is a BPS state.

The present analysis can be extended in various directions. First
of all it would be very interesting and challenging to find the
boundary theory analogue of the present study. It is not clear
whether the theory is integrable from the boundary theory. But
given the result of the present paper it is interesting to find
out operators in the LST that correspond to the giant magnons in
the NS5-brane near horizon geometry. Further one can try to study
the connection of this particular giant magnon solution to the
pp-wave of the NS5-brane background, which is the Nappi-Witten
model in the lines of \cite{Maldacena:2006rv}.

\vskip .5in \noindent {\bf Acknowledgements:} We would like to
thank S. Minwalla, J. Govaerts and W. H. Huang for useful
correspondence. RRN would like to thank Bum-Hoon Lee and Sunggeun
Lee for interesting discussions. The work of JK was supported in
part by the Czech Ministry of Education under Contract No. MSM
0021622409, by INFN, by the MIUR-COFIN contract 2003-023852, by
the EU contracts MRTN-CT-2004-503369 and MRTN-CT-2004-512194, by
the INTAS contract 03-516346 and by the NATO grant PST.CLG.978785.
The work of RRN was supported by the Science Research Center
Program of the Korean Science and Engineering Foundation through
the Center for Quantum Spacetime (CQUeST) of Sogang University
grant no. R11-2005-021.

\end{document}